\documentclass{article}
\usepackage[latin9]{inputenc}
\usepackage{amsmath}
\usepackage{graphicx}

\makeatletter

\newcommand{\noun}[1]{\textsc{#1}}

\usepackage{amsfonts}

\makeatother

\begin{document}

\title{An investigation into the source of stability of the electron spin
projections}

\author{A. M. Cetto and L. de la Peña \\
Instituto de Física, Universidad Nacional Autónoma de México\\
 Ciudad de México, Mexico}
\maketitle
\begin{abstract}
We propose that the stability of the two projections of the electron
spin along the direction of an applied magnetic field, is an effect
of high-frequency vibrations acting on the spin magnetic moment. The
source of the high-frequency vibrations is to be found in the zero-point
radiation field. 
\end{abstract}

\section{Introduction}

In contrast with a classical compass needle, which points invariably
towards the North pole, in the presence of an external magnetic field
the spin of an electron orients itself either parallel or antiparallel
to the direction of the field. This is such a well-established fact
in quantum theory that nobody ever seems to wonder about it, let alone
try to find an explanation for it. The double sign of the spin projection
is taken for granted, even though the parallel orientation appears
as classically counterintuitive, as it is energetically less favourable
than any other orientation of the spin with respect to the field:
it corresponds in the classical case to a position of unstable equilibrium.

Yet also in classical mechanics there exist systems that can have
two positions of stable equilibrium in opposite directions. The best
known instance of such a system is the inverted pendulum, which is
used as a benchmark in control theory and finds numerous technical
applications, some as popular as the self-balancing scooter. Without
going so far, we refer to the experience of balancing a broom that
stands upside down on the palm of our hand, by making small, rapid
movements with the hand to keep it upright. Such remarkable state
of motion is the result of adding to the pendulum a vibrating motion
of high frequency applied to the supporting point \cite{1 Thom2}.
This would suggest the idea that something similar can be occurring
in the case of the spin magnetic moment. But such an idea would seem
not to work at all, if no high-frequency source is available.

Here we address this enigma by appealing to the existence of the vacuum
or zero-point radiation field, as suggested by stochastic electrodynamics
(\textsc{sed}). In this theory, proposed as a foundation for quantum
mechanics, a central role is played by the zero-point radiation field
(\textsc{zpf}), taken as a Maxwellian field that fills the whole space
and covers the entire frequency spectrum (see \emph{e.g}. \cite{2 TEQ}
and references therein). In particular, the high-frequency modes (of
Compton's frequency) acting on the electron have been shown elsewhere
to be related to the origin of de Broglie's wave \cite{3 Dice} and
to (a nonrelativistic version of) the \emph{zitterbewegung} \cite{4 Milonni}.

It is therefore natural within \textsc{sed} to consider that also
the magnetic moment of the electron is subject to the action of these
\textsc{zpf} modes. This is what we do in the present paper, with
the purpose of finding an explanation for the two positions of stable
equilibrium of the (quantum) electron spin. In section \emph{\ref{Compton}}
we present the model of the electron spin as it emerges from \textsc{sed},
to be used for this purpose. In section 3 we establish the equations
of motion for the electron subject to an applied magnetic field in
addition to the \textsc{zpf}, and solve them for the one-dimensional
case by separating the fast variables from the slow ones, which leads
to a direct demonstration of the stability of the two spin states,
as seen in section 4. The extra vibrations of the magnetic moment
of the electron due to its interaction with the high-frequency magnetic
component of the \textsc{zpf} result in an effective potential with
two deep minima. As shown in section 5, under conditions that hold
in real experiments these minima correspond to the two positions of
stable equilibrium of the spin magnetic moment: spin 'up' and spin
'down'. In section 6 the same conclusion is seen to apply in the two-dimensional
case, that takes into account the Larmor precession. The paper concludes
with a brief recapitulation.

\section{The spin and the zero-point field\label{Compton}}

Before proceeding with the calculations let us briefly present the
model of the electron spin to be used in this paper. In quantum theory
the electron spin is represented with the aid of the Pauli matrices,
which automatically introduce all the required quantum properties.
The situation is quite different in \textsc{sed}, a theory based on
the hypothesis that quantum mechanics is the result of the action
of the radiation \textsc{zpf} on an otherwise classical system. This
theory has evolved in the course of time to reach a well-developed
status (see \emph{e.g.} ref. \cite{2 TEQ}) that includes, in particular,
a theory of the electron spin, just the one we take here as the basis
for our proposal.

The most immediate effect of the presence of the \textsc{zpf} is that
an electron, which normally is part of an atomic system, acquires
a stochastic motion. The \textsc{zpf} covers the entire spectrum;
but not all frequencies have the same importance for the atomic system.
As is well known, Dirac's equation for the free electron reveals the
existence of the \emph{zitterbewegung}, a helicoidal motion of frequency
of the order of $\omega_{C}$ and amplitude of the order of $\lambda_{C},$
where $\omega_{C}=mc^{2}/\hbar$ represents the Compton frequency
of the electron of mass\ $m$, and $\lambda_{C}$ the corresponding
wavelength $\lambda_{C}=h/mc.$ In the textbooks on quantum mechanics,
$\lambda_{C}$ appears normally only in association with the Compton
effect and related considerations. When looking beyond the quantum
formalism, $\lambda_{C}$ reappears, with an important though indirect
role as the source of de Broglie's wavelength (\cite{2 TEQ}, \cite{3 Dice}).
In \textsc{sed} \textemdash as in nonrelativistic \textsc{qed}\textemdash{}
$\omega_{C}$ is used frequently as a convenient cutoff to regularize
integrals that determine \emph{e.g.} the average properties of dynamical
variables. This leads to the emergence of particle oscillations of
frequencies of order $\omega_{C}$. The introduction of the cutoff
is thus of physical significance rather than a simple mathematical
device, meaning that the electron (as all matter) becomes transparent
to the radiation field of very high frequencies.

The physical meaning assigned to the cutoff leads to an also physically
meaningful finite effective size for the electron, of the order of
$\lambda_{C}$. Such effective size has only a statistical (or coarse-grained)
sense, since it is the result of the fluctuations or rapid oscillations
impressed upon the particle by the field, the original electron being
still a point particle; it is the interaction with the field that
dresses it. This view coincides with the \textsc{qed} picture, where
the electron acquires through its electromagnetic interactions an
effective size of the order of $\lambda_{C}$ (see \emph{e.g.} \cite{4 Milonni}).

We thus arrive at a consistent picture of the electron as a small
sphere of effective radius of order $\lambda_{C}$ realizing a kind
of nonrelativistic \emph{zitterbewegung} of frequency of order $\omega_{C}$.
As a consequence of the torque exerted by the electric field modes
of a given circular polarization, the electron describes a helicoidal
motion, which results in a mean intrinsic angular momentum of value
$\hbar/2$, mean square angular momentum $3\hbar^{2}/4,$ and an associated
magnetic moment with $g$-factor of 2. The spin is thus identified
as an emergent property generated by the action of the \textsc{zpf},
as shown in detail in refs. \cite{5 CePeVal14}-\cite{7 CePeVal17}.

The approach used in \textsc{sed} may seem (to some) too classical
to reproduce the quantum properties of matter. However, as shown \emph{e.g.}
in chapter 5 of ref. \cite{2 TEQ}, when the \textsc{sed} system transits
to a state in which it has acquired ergodic properties (and also energy
balance holds), the appropriate description of the dynamics becomes
one in which the variables are represented by matrices and the whole
Hilbert-space formalism of quantum mechanics applies \textemdash including
the electron spin with its two projections. In brief, a \textit{qualitative}
change of state occurs in the behaviour of the system (as a sort of
phase transition), that demands a leap in its description from classical
to quantum. The present paper is intended to shed some light on the
physical mechanism leading precisely to the two spin projections.

Such abrupt changes in the description are not unknown to physics.
A traditional example is the transition (in both, classical and quantum
physics) in the description of a system of a few particles to a system
with a huge number of them, which requires a statistical treatment.
A more recent example is that of nonlinear dissipative systems, in
which the phenomenon of deterministic chaos characteristically takes
place, contrary in principle to the traditional classical behaviour.
In both cases new notions become indispensable in replacement of the
older ones.

\section{The effective potential}

Let us consider an electron with its spin and magnetic moment $\boldsymbol{\mu}$
forming an angle $\theta$ with the $z$-axis; the applied magnetic
field is $\boldsymbol{B=}\boldsymbol{\hat{e}}_{z}B,$ with $B$ constant
for simplicity. The corresponding interaction energy is $V=-\boldsymbol{\mu\cdot B}.$
Since $\mu=e\hbar/2mc=-\mu_{0}<0,$ the energetically stable orientation
of the spin is opposite to the direction of the magnetic field. However,
quantum theory considers also the direction along the field to correspond
to a stable solution, as has been experimentally proven again and
again.

With the purpose of finding an explanation for the stability of the
spin orientation in both directions, we take into account that the
above system is subject to the action of the \textsc{zpf} of high
frequency. For our present purposes it is enough to consider a mode
of this field of a sufficiently high frequency $\gamma$, uniformly
distributed on the $xy$-plane and directed along the $z$-axis; we
therefore write the magnetic component of this mode as $\boldsymbol{B}_{0}(t)=\boldsymbol{\hat{e}}_{z}B_{0}\cos\gamma t.$
In spherical coordinates, the Lagrangian for the problem is 
\begin{equation}
L=\frac{1}{2}I\left(\dot{\theta^{2}}+\dot{\varphi^{2}}\sin^{2}\theta\right)-\mu_{0}(B\cos\theta+B_{0}\cos\theta\cos\gamma t),\label{3.2}
\end{equation}
with $I$ an effective moment of inertia to be determined below. The
corresponding equations of motion are 
\begin{equation}
\ddot{\theta}-\dot{\varphi^{2}}\sin\theta\cos\theta=\frac{\mu_{0}B}{I}\sin\theta+\frac{\mu_{0}B_{0}}{I}\sin\theta\cos\gamma t,\label{3.4}
\end{equation}
\begin{equation}
\dot{\varphi}\sin^{2}\theta=\text{constant}.\label{3.6}
\end{equation}
Note that the second term on the l.h.s. is a kinematic term due to
the use of spherical variables; it does not disappear in the absence
of the magnetic field. Further, the variable $\varphi$ is ignorable,
and the angular velocity about the $z$-axis, given by $\dot{\varphi}$,
is arbitrary. Let us, for clarity, assume in a first instance that
initially $\dot{\varphi}=0$ and $\varphi=0$; then according to eq.
(\ref{3.6}) the magnetic moment vector remains on the $xz$-plane,
and (\ref{3.4}) reduces to 
\begin{equation}
\ddot{\theta}=\frac{\mu_{0}B}{I}\sin\theta+\frac{\mu_{0}B_{0}}{I}\sin\theta\cos\gamma t.\label{3.8}
\end{equation}
Now we follow the usual procedure \cite{1 Thom2} to determine the
effect of the high-frequency term on the orientation of $\boldsymbol{\mu}$,
by separating the fast terms from the slowly varying component of
the motion. This procedure applies when $\gamma\gg\omega_{0}$, where
$\omega_{0}$ is a frequency parameter associated with the external
field, 
\begin{equation}
\omega_{0}\equiv\sqrt{\frac{\mu_{0}B}{I}}.\label{m15}
\end{equation}
Under this condition, a first-order determination of the effects of
the high-frequency vibration is sufficient, as will become clear below
(see also \cite{1 Thom2}, \cite{8 LanLif}). Thus we write $\theta=\Theta+\xi,$
with $\Theta$ the slow (dominant) motion component and $\xi$ the
fast (small, high-frequency) correction. A Taylor series expansion
of eq. (\ref{3.8}) gives to first order in $\xi$ 
\begin{equation}
\ddot{\Theta}+\ddot{\xi}=\omega_{0}^{2}\left(\sin\Theta+\xi\cos\Theta\right)+\frac{\omega_{0}^{2}B_{0}}{B}\left(\sin\Theta\cos\gamma t+\xi\cos\Theta\cos\gamma t\right).\label{m16}
\end{equation}
This equation contains both the smooth and the rapidly vibrating terms,
which must be separately equal. For the latter we get 
\begin{equation}
\ddot{\xi}=\frac{\omega_{0}^{2}B_{0}}{B}\sin\Theta\cos\gamma t+\mathcal{O}(\xi),\label{m17}
\end{equation}
which gives to first order 
\begin{equation}
\xi=-\frac{\omega_{0}^{2}B_{0}}{\gamma^{2}B}\sin\Theta\cos\gamma t.\label{m18}
\end{equation}
Substituting this in (\ref{m16}) and averaging over the short period
$2\pi/\gamma,$ an interval during which $\Theta$ remains virtually
the same, we obtain an equation for the slow motion, 
\begin{eqnarray}
\ddot{\Theta} & = & \omega_{0}^{2}\sin\Theta-2\Omega^{2}\frac{B}{B_{0}}\sin\Theta\cos\Theta\overline{\cos\gamma t}^{t}+\frac{\omega_{0}^{2}B_{0}}{B}\sin\Theta\overline{\cos\gamma t}^{t}\nonumber \\
 &  & -2\Omega^{2}\sin\Theta\cos\Theta\overline{\cos^{2}\gamma t}^{t},\label{m20}
\end{eqnarray}
which simplifies into 
\begin{equation}
\ddot{\Theta}=\omega_{0}^{2}\sin\Theta-\Omega^{2}\sin\Theta\cos\Theta,\label{m22}
\end{equation}
with the frequency parameter $\Omega$ defined as 
\begin{equation}
\Omega\equiv\frac{\mu_{0}B_{0}}{\sqrt{2}I\gamma}.\label{ss17}
\end{equation}

Since in what follows we are interested in the slow motion only, described
by eq. (\ref{m22}), we shall go back to the original notation for
the angle, \emph{i.e.}, $\Theta\rightarrow\theta$. Equating now the
angular acceleration $I\ddot{\theta}$ to (minus) the derivative of
an effective potential energy $V_{\text{eff}}$ associated with the
magnetic moment, we obtain 
\begin{equation}
V_{\text{eff}}=I\omega_{0}^{2}\cos\theta+\frac{1}{2}I\Omega^{2}\sin^{2}\theta.\label{ss10}
\end{equation}
This potential is represented in fig. 1, for two values of $a/2=\omega_{0}^{2}/\Omega^{2}<1$
(see (\ref{ss19}) below).
\begin{figure}
\begin{centering}
\includegraphics[scale=0.5]{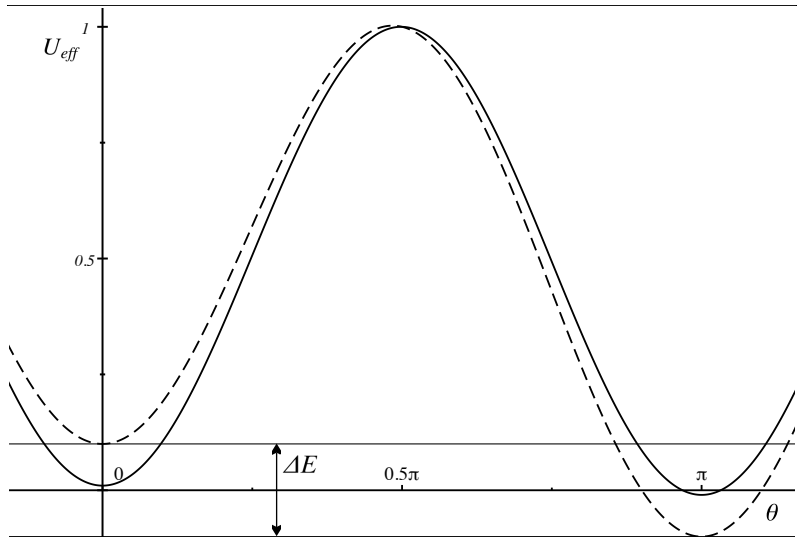}
\par\end{centering}
\caption{Effective potential $U_{eff}=(2/I\Omega^{2})V_{eff}=a\cos\theta+\sin^{2}\theta$,
for $a=10^{-2}$ (solid line) and $a=10^{-1}$(dashed line). Current
experimental values are $a\ll10^{-6}$. The energy gap is $\varDelta E=2a$.}
\end{figure}

\section{Stable equilibrium positions}

As is well known from the theory of high-frequency excitations (\cite{1 Thom2}),
a first-order consequence of these excitations is an additional (mean)
force applied on the system, able to produce peculiar effects. Here
the additional force derives from the second term in $V_{\text{eff}}$
(see eq. (\ref{m22})), and the effect of it is the emergence, under
appropriate conditions, of two positions of stable equilibrium for
$\boldsymbol{\mu}$ at the minima of the potential curve represented
in fig. 1. This situation is closely analogous to the behaviour of
the inverted pendulum (see \emph{e.g.} \cite{1 Thom2}), \cite{8 LanLif},
\cite{9 Butikov}). Indeed, a comparison of these two theories shows
that eq. (\ref{m22}) is common to both, with the appropriate parameters
and dynamical variables in each case.

To determine the stable equilibrium positions we look for the solutions
of the equation 
\begin{equation}
\frac{dV_{\text{eff}}}{d\theta}=0\quad\text{or }\quad\left(-\omega_{0}^{2}+\Omega^{2}\cos\theta\right)\sin\theta=0,\label{ss12}
\end{equation}
whence both $\theta=0,\pi$ are equilibrium solutions. Since

\begin{equation}
\frac{d^{2}V_{\text{eff}}}{d\theta^{2}}=I\left(-\omega_{0}^{2}\cos\theta+\Omega^{2}\left(\cos^{2}\theta-\sin^{2}\theta\right)\right),\label{ss14}
\end{equation}
it is clear from the outset that $\theta=\pi$ is a stable solution
for any value of the parameters. This represents the more energetically
favourable equilibrium position (the only stable one in the classical
case). For $\theta=0$ the second derivative $d^{2}V_{\text{eff}}/d\theta^{2}$
has positive values if and only if 
\begin{equation}
\omega_{0}^{2}<\Omega^{2}.\label{ss19}
\end{equation}
When this condition is satisfied, also the solution $\theta=\pi$
is stable. In terms of the magnetic fields, (\ref{ss19}) rewrites
as

\begin{equation}
B<\frac{\mu_{0}B_{0}^{2}}{2I\gamma^{2}},\label{ss16}
\end{equation}
which means that for a very strong applied magnetic field $B$ the
stability of this solution may be lost.

To estimate the frequencies $\omega_{\pm}$ of oscillation around
the equilibrium points we make the usual small-amplitude approximation
\begin{equation}
\left.\frac{d^{2}V_{\text{eff}}}{d\theta^{2}}\right\vert _{\theta=0,\pi}=I\omega_{\pm}^{2},\label{ss20}
\end{equation}
with the plus sign corresponding to $\theta=\pi$. Combining with
eq. (\ref{ss14}) and using (\ref{m15}) and (\ref{ss17}), this gives
\begin{equation}
\omega_{\pm}^{2}=\Omega^{2}\pm\omega_{0}^{2}.\label{ss22}
\end{equation}
It is clear from fig. 1 that for $\omega_{0}^{2}\ll\Omega^{2}$, the
two equilibrium positions are virtually equally stable. The difference
between the two frequencies is given in this case, according to eq.
(\ref{ss22}), by 
\begin{equation}
\Delta\omega=\omega_{+}-\omega_{-}=\Omega\left[\sqrt{1+\left(\frac{\omega_{0}}{\Omega}\right)^{2}}-\sqrt{1-\left(\frac{\omega_{0}}{\Omega}\right)^{2}}\right]\simeq\frac{\omega_{0}^{2}}{\Omega},\label{ss24}
\end{equation}
the more energetically favourable one being slightly higher.

Further, notice that the energy difference between the two stable
solutions is determined solely by the value of the applied field,
\begin{equation}
\Delta E=2\mu_{0}B,\label{ss18}
\end{equation}
with independence of the values of the parameters associated with
the \textsc{zpf}.

\section{Stability of the electron spin projections\label{values}}

In order to apply the above results to the specific case of the electron
spin we must assign values to the various physical parameters involved.
According to the discussion in section \ref{Compton}, we take $\omega_{C}$
as the frequency of the vibrating mode of the \textsc{zpf} acting
on the spin magnetic moment. This gives for the moment of inertia,
defined as $I\simeq\hbar/\gamma$ for an angular momentum of the order
of $\hbar,$ the (approximate) value 
\begin{equation}
I\simeq\frac{\hbar}{\omega_{C}}=\frac{\hbar^{2}}{mc^{2}}.\label{e25}
\end{equation}
By writing the moment of inertia in terms of the mass and the effective
radius of the electron acquired as a result of the \emph{zitterbewegung},
$I\simeq mr_{eff}^{2},$ we obtain for the latter the following value:
\begin{equation}
r_{eff}\simeq\sqrt{\frac{I}{m}}\simeq\frac{\hbar}{mc}=\frac{\lambda_{C}}{2\pi},\label{ss26}
\end{equation}
\emph{i.e.}, the effective size of the electron turns out to be of
the order of Compton's wavelength, in agreement with the standard
literature (see \emph{e.g}. ref. \cite{4 Milonni}) and our previous
discussion.

We need also an estimate for the magnetic field amplitude $B_{0}$,
which is related to the energy density of the \textsc{zpf} mode of
Compton's frequency. This energy density is given by the total energy
of this mode, $(1/2)\hbar\omega_{C}$, divided by the effective volume
occupied by the electron, $V_{eff}=(4/3)\pi r_{eff}^{3}$, which gives
\begin{equation}
B_{0}^{2}=\frac{3}{2}(\frac{2\pi}{\lambda_{C}})^{3}\hbar\omega_{C},\label{27}
\end{equation}
whence we obtain from eq. (\ref{ss17}), with $\mu_{0}=\mid e\mid\hbar/2mc$,
\begin{equation}
\Omega^{2}=\frac{3}{8}\alpha\omega_{C}^{2},\label{ss27'}
\end{equation}
where $\alpha=e^{2}/\hbar c$ is the fine-structure constant. With
these order-of-magnitude estimates the condition (\ref{ss19}) for
the existence of two stable equilibrium positions reads 
\begin{equation}
\omega_{0}^{2}<\frac{3}{8}\alpha\omega_{C}^{2},\label{ss28}
\end{equation}
or in terms of the magnitude of the applied field $B$, according
to eq. (\ref{m15}), 
\begin{equation}
\mu_{0}B<\frac{3}{8}\alpha\hbar\omega_{C}=\frac{3}{8}\alpha mc^{2}.\label{ss29}
\end{equation}
For a comparison with experiment it is useful to rewrite (\ref{ss29})
as a condition on the Larmor frequency, 
\begin{equation}
\omega_{L}=\frac{\mu_{0}B}{\hbar}<\frac{3}{8}\alpha\omega_{C}.\label{ss29'}
\end{equation}
Electron paramagnetic resonance (EPR) measurements are usually carried
out with microwaves of frequencies of the order of $\omega_{L}\simeq10^{10}$
s$^{-1}$, whilst the Compton frequency of the electron is $\omega_{C}\simeq10^{21}$
s$^{-1}$ . Such Larmor frequencies correspond to magnetic field strengths
of the order of 0.35 T. Magnetic fields of extremely high intensity,
of the order of 10 T, correspond to Larmor frequencies close to $\omega_{L}\simeq10^{13}s^{-1},$
still eight orders of magnitude smaller than $\omega_{C}$. Therefore,
under present experimental situations the stability condition (\ref{ss19})
is amply satisfied. This means that the stability of the two equilibrium
positions $\theta=0,\pi$ is well guaranteed for a wide range of values
of the applied magnetic field.

Further, note that according to eq. (\ref{ss24}), the relative difference
between the two frequencies of oscillation around the points of equilibrium
is of the order of

\begin{equation}
\frac{\Delta\omega}{\Omega}\simeq(\frac{\omega_{0}}{\Omega})^{2}=\frac{8}{3\alpha}\frac{\omega_{L}}{\omega_{C}},\label{ss30}
\end{equation}
which represents an insignificant deviation for usual magnetic field
strengths. fig. 1 shows that the (near harmonic) potential wells are
indeed very similar in width, and deep enough to sustain the stability
of both solutions.

\section{Inclusion of the Larmor precession}

It is possible to extend the analysis carried out above to the more
general case in which the magnetic moment rotates in two dimensions.
This allows us to take into account the effect of the torque exerted
on the magnetic moment by the magnetic force, 
\begin{equation}
\boldsymbol{\tau}=\boldsymbol{\mu\times B}=\frac{d\boldsymbol{\mu}}{dt}.\label{2d12}
\end{equation}
As is well known from classical electrodynamics, this torque gives
rise to the Larmor precession, which is a rotation of $\boldsymbol{\mu}$
about the $z$-axis, with angular velocity given by $\dot{\varphi}=\omega_{L}=\mu_{0}B/\hbar$
(see eq. (\ref{ss29'})), independent of the zenithal angle $\theta$.
Since this movement of precession is orthogonal to the motion along
the $xz$-plane described above (with $\theta$ as the only variable),
and $\varphi$ is not a rapidly oscillating variable, the stability
behaviour should be essentially the same as above. To show that this
is the case, we rewrite the complete equation (\ref{3.4}) in terms
of the Larmor frequency for $\dot{\varphi}$, 
\begin{equation}
\ddot{\theta}=\omega_{0}^{2}\sin\theta+\omega_{L}^{2}\sin\theta\cos\theta+\frac{\omega_{0}^{2}B_{0}}{B}\sin\theta\cos\gamma.\label{2d18}
\end{equation}
By separating the slow motion from the fast terms in this equation,
in analogy with the 1-D case, one obtains for the effective potential,
using once more (\ref{ss17}), 
\begin{equation}
V_{eff}=I\omega_{0}^{2}\cos\theta+\frac{I}{2}\left(\Omega^{2}-\omega_{L}^{2}\right)\sin^{2}\theta.\label{2d20-1}
\end{equation}
As before, both values $\theta=0,\pi$ correspond to points of stability.
From 
\begin{equation}
\frac{d^{2}V_{\text{eff}}}{d\theta^{2}}=-I\omega_{0}^{2}\cos\theta+I\left(\Omega^{2}-\omega_{L}^{2}\right)\left(\cos^{2}\theta-\sin^{2}\theta\right)\label{2d21}
\end{equation}
we see that for 
\begin{equation}
\omega_{0}^{2}\left(1+\frac{\omega_{L}^{2}}{\omega_{0}^{2}}\right)<\Omega^{2},\label{2d22}
\end{equation}
again both equilibrium positions are stable. Since, for the values
of the parameters given in section \ref{values} (see the discussion
following eq. (\ref{ss29'})), 
\begin{equation}
\frac{\omega_{L}^{2}}{\omega_{0}^{2}}=\frac{\mu_{0}B}{\hbar\omega_{C}}\ll1,\label{2d24}
\end{equation}
essentially the same stability condition (\ref{ss19}) holds in the
more general case that includes the Larmor precession.

\section{Conclusions}

Stochastic electrodynamics shows us that an electron spin free to
rotate under the combined action of an applied field $\boldsymbol{B}$
and the high-frequency oscillatory mode $\boldsymbol{B}_{0}\cos\gamma t$
of the \textsc{zpf} parallel to $\boldsymbol{B},$ has two stable
equilibrium positions, parallel and antiparallel to $\boldsymbol{B}$,
just as described in quantum mechanics. The condition for these two
spin projections, expressed in (\ref{ss19}), is largely satisfied
for present experimental values of the magnetic field intensity.

\end{document}